\documentclass[preprint,12pt]{elsarticle}

\usepackage{amssymb}
\usepackage{amsmath}
\usepackage{amsmath,amsfonts,amssymb}
\usepackage{graphicx}
\usepackage{url}
\usepackage{booktabs}
\usepackage{longtable}
\usepackage{multirow}
\usepackage{hyperref}
\usepackage{float}
\usepackage{tabularx}
\usepackage{array}
\usepackage{tabularx}
\usepackage{ragged2e}

\usepackage{graphicx}
\usepackage{algorithm}
\usepackage{float}
\usepackage{listings}
\usepackage{tcolorbox}
\usepackage{float}
\usepackage{tabularx}
\usepackage{enumitem}
\usepackage{booktabs}
\usepackage{enumitem}   
\usepackage{caption}    
\usepackage{array}      

\usepackage{booktabs}
\usepackage{tabularx}
\usepackage{makecell}
\usepackage{array}
\usepackage{float}

\newcolumntype{C}{>{\centering\arraybackslash}X}
\usepackage{placeins}  

\journal{Information Sciences}

\begin{document}

\begin{frontmatter}



\title{ExAI5G: A Logic-Based Explainable AI Framework for Intrusion Detection in 5G Networks}


\author[1]{Saeid Sheikhi}
\author[1]{Panos Kostakos}
\author[1]{Lauri Loven}
\affiliation[1]{organization={Faculty of Information Technology and Electrical Engineering,\\ University of Oulu},
            addressline={},
            city={Oulu},
            postcode={90570},
            state={},
            country={Finland}}

\begin{abstract}
Intrusion detection systems (IDSs) for 5G networks must handle complex, high-volume traffic. Although opaque "black-box" models can achieve high accuracy, their lack of transparency hinders trust and effective operational response. We propose \emph{ExAI5G}, a framework that prioritizes interpretability by integrating a Transformer-based deep learning IDS with logic-based explainable AI (XAI) techniques. The framework uses Integrated Gradients to attribute feature importance and extracts a surrogate decision tree to derive logical rules. We introduce a novel evaluation methodology for LLM-generated explanations, using a powerful evaluator LLM to assess \textbf{actionability} and measuring their \textbf{semantic similarity} and \textbf{faithfulness}. On a 5G IoT intrusion dataset, our system achieves \textbf{99.9\%} accuracy and a \textbf{0.854} macro F1-score, demonstrating strong performance. More importantly, we extract 16 logical rules with \textbf{99.7\%} fidelity, making the model's reasoning transparent. The evaluation demonstrates that modern LLMs can generate explanations that are both faithful and actionable, indicating that it is possible to build a trustworthy and effective IDS without compromising performance for the sake of marginal gains from an opaque model.
\end{abstract}



\end{frontmatter}



\section{Introduction}
The rollout of 5G networks has enabled massive connectivity for IoT devices and critical services, but it also expands the attack surface for cyber intrusions \cite{radoglou20235gcids,fan2020iotdefender}. Machine learning–based Intrusion Detection Systems (IDS) are being adopted to identify anomalous traffic \cite{fan2020iotdefender,sheikhi2023ddos}. While deep learning models can achieve high detection rates, their decisions are often unclear, creating a trust gap for security operators\cite{radoglou20235gcids, linkov2020cybertrust}. This raises an important question for security-critical domains: a slight improvement in performance metrics may not be worth the risk of an unexplainable system.

Prior work has emphasized explainability in cybersecurity, for instance, by extracting decision rules or visualizing feature importance\cite{subasi2024critical}. However, many XAI methods focus on feature-importance explanations (e.g., SHAP or LIME) that can be unstable and difficult to translate into actionable insights\cite{subasi2024critical,charmet2022explainable}. There is a need for \emph{logic-based} explanations, unambiguous rules defining the conditions under which an alert is triggered, to support expert reasoning and verification.

In this paper, we introduce \textbf{ExAI5G}, an explainable AI framework that directly addresses the performance-vs-interpretability trade-off. The proposed approach first trains a Transformer-based IDS model, then makes it transparent by applying Integrated Gradients\cite{sundararajan2017axiomatic} to quantify feature importance and training a surrogate decision tree to extract human-readable logical rules. Finally, we leverage multiple LLMs to convert these rules and attributions into natural language and introduce a new methodology to quantitatively evaluate their quality, faithfulness, and actionability.

The main contributions of the paper are as follows:
\begin{enumerate}
    \item A novel XAI framework that makes a deep learning IDS transparent via logic rule extraction and validated natural language explanations.
    \item A new three-part validation scheme to assess LLM-generated explanations, measuring semantic similarity, attribution faithfulness, and an expert-level actionability score.
    \item A direct comparison with opaque, high-performing models, demonstrating that our explainable system achieves strong performance (99.9\% accuracy) with a compact, high-fidelity (99.7\%) rule set.
    \item A quantitative analysis of explanations from four different LLMs, proving that modern models can achieve perfect scores in faithfulness and actionability.
    \item A pilot expert study with two security analysts, comparing human ratings of explanation quality to the LLM-as-judge metrics and highlighting where they align and diverge.
\end{enumerate}

ExAI5G aims to enhance the transparency and reliability of AI-driven security systems by prioritizing trustworthiness alongside performance. The rest of the paper is organized as follows: Section 2 reviews related work; Section 3 presents our methodology; Section 4 outlines the experimental setup; Section 5 reports the results; Section 6 discusses our findings. Finally, Section 7 concludes the paper.

\section{Related Work}

\textbf{Intrusion Detection in 5G/IoT Networks.} The security of 5G core and IoT networks has been the focus of extensive research, with various IDS solutions proposed to handle novel attack vectors and the scale of 5G traffic. Traditional signature-based methods struggle with new or evolving threats, leading to a surge in anomaly-based IDS using machine learning. For example, Fan \emph{et al.} introduced \emph{IoTDefender}, a federated transfer learning framework for 5G IoT intrusion detection that aggregates models from edge devices to improve detection of attacks across distributed data\cite{fan2020iotdefender}. Sood \emph{et al.} proposed an anomaly detection scheme for 5G networks using dimensionality reduction to preprocess features, improving classification efficiency for attacks such as unauthorized access and Denial of Service (DoS)\cite{sood2023intrusion}. In 5G contexts, Kim \emph{et al.} focused on effective feature selection to identify Distributed Denial-of-Service (DDoS) attacks in a 5G core network environment, highlighting the importance of choosing discriminative features to handle high-volume IoT traffic\cite{kim2022effective}. These works demonstrate high detection rates but essentially treat the ML models as black boxes. As 5G IDS deployments become more complex (e.g., deep neural networks, federated learning), understanding model decisions becomes crucial for debugging and compliance.

\textbf{Explainable AI in Cybersecurity.} Explainable AI has been utilized in security domains to build trust in automated decision-making processes. A recent survey by Charmet \emph{et al.} reviews XAI techniques for cybersecurity, reporting that most approaches either visualize feature importances or provide example-based explanations (prototypes, counterfactuals) rather than logical reasoning\cite{charmet2022explainable}. They emphasize the need for explanations that security analysts can act on, aligning with Linkov \emph{et al.}’s concept of moving “from explainable to actionable” AI\cite{linkov2020cybertrust}. In intrusion detection, many studies employ post-hoc explanation methods such as Local Interpretable Model-agnostic Explanations (LIME) or Shapley Additive exPlanations (SHAP) to interpret deep learning models\cite{nyre2022explainable}. Gaspar \emph{et al.} (2024) emphasize the challenges posed by LIME and SHAP in cybersecurity, indicating that the instability of feature importance ratings can hinder the trust and usability of these approaches for security practitioners. These methods can facilitate a better understanding of model decisions. However, their fluctuating outputs when applied to similar datasets raise doubts about their reliability \cite{gaspar2024explainable}. More interpretable-by-design models like decision trees or rule-based classifiers have been revisited for IDS to offer transparency. Gyawali \emph{et al.} integrated an explainability module into an IoT anomaly detection system, showing that highlighting feature importance (e.g., via heatmaps) helped administrators grasp why an alert was raised\cite{gyawali2024leveraging}. Similarly, Siganos \emph{et al.} proposed an explainable AI–based IDS for IoT, combining deep learning with an explanation interface to present the reasons for detections (such as particular network features being outside normal ranges)\cite{siganos2023explainable}. Our work builds on this literature by providing not just feature importance but also logical if-then rules that succinctly characterize attack traffic versus benign traffic, which can be more actionable (e.g., as firewall rules or forensic insights) than raw feature weights.

\textbf{Logic-Based Rule Extraction and Verification.} Using logic to interpret ML models has a rich history. Early work by Craven \& Shavlik introduced methods like TREPAN for extracting decision trees from trained neural networks, aiming to approximate the network’s decisions with a set of logical conditions\cite{craven1995extracting}. Similarly, rule extraction algorithms such as DeepRED (Zilke \emph{et al.}, 2016) decompose a deep neural network into equivalent rule sets\cite{zilke2016deepred}. These approaches ensure that each explanation (rule) corresponds to a region in feature space with a consistent predicted class, offering global insight into the model. In security, rule-based systems have long been used (e.g., Snort signatures), so being able to convert a learned IDS into rules helps in bridging data-driven models with expert systems. Recent studies have enhanced rule extraction with probabilistic reasoning; for instance, Contreras \emph{et al.} combined logic rules with embedding analysis to explain deep models, producing rules that capture feature interactions in a comprehensible manner\cite{contreras2024explanation}. Logic-based explanations can also be formally verified or checked against domain knowledge (for example, verifying that a rule for detecting port scan attacks aligns with known indicator-of-compromise patterns). In our Framework, we use a surrogate decision tree (depth-limited) to extract rules that describe the IDS model’s behavior. This not only provides an interpretable global model but also facilitates \emph{logic verification}: we can examine if the extracted rules make sense (e.g., an IoT DoS attack rule might involve a high rate of MQTT messages) and whether any rules conflict or are redundant. Assessing the fidelity of the rule set to the original model ensures that the logical abstraction remains accurate.

\textbf{LLMs for Explainable AI.} Large Language Models have recently been explored as tools for enhancing XAI by generating human-readable explanations from model data\cite{bilal2025llms}. The conversational and reasoning abilities of LLMs (e.g., GPT-3.5, GPT-4) allow them to take structured information (like a set of rules or a feature attribution list) and produce a coherent narrative explanation\cite{guidotti2018survey}. A comprehensive survey by Bilal \emph{et al.} discusses how LLMs can serve as intermediaries between complex model outputs and user-friendly explanations, highlighting use cases in which LLMs translate model decisions into domain language\cite{bilal2025llms}.

\section{Methodology}
\begin{figure*}[h]
\centering
\includegraphics[width=\textwidth]{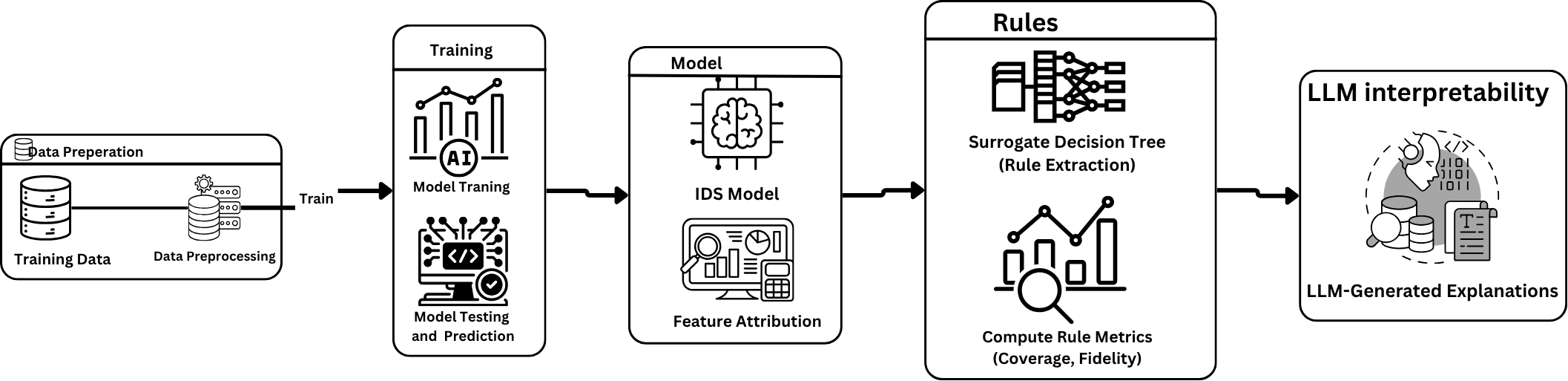}
\caption{ExAI5G Framework: a TabTransformer is trained on 5G network data; Integrated Gradients highlight important features; a surrogate decision tree produces rules; an LLM produces a high-level explanation.}
\label{fig:pipeline}
\end{figure*}

The ExAI5G framework comprises four main stages: (1) a \textbf{Transformer IDS model} that learns to detect intrusions; (2) an \textbf{Integrated Gradients (IG) attribution} mechanism to evaluate feature importance; (3) a \textbf{decision tree surrogate} model to approximate the Transformer’s decision function, from which we \textbf{extract logical rules}; and (4) an \textbf{LLM-mediated explanation module} that generates and precisely validates natural language descriptions of the model’s behavior. Figure~\ref{fig:pipeline} illustrates the framework at a conceptual level.

\noindent\textbf{Data Splitting and Preprocessing:}
To prevent data leakage and ensure a robust evaluation, we first partition the raw training data into an 80\% training set and a 20\% validation set using stratified sampling. A \texttt{StandardScaler} is fitted \textbf{only on the new, smaller training set}. This same fitted scaler is then used to transform the training, validation, and test sets, ensuring that no information from the validation or test data influences the training process.

\noindent\textbf{Transformer IDS Model:} 
We first transform all categorical and string-valued fields into numeric representations and then standardize the features. Our Transformer-like network projects each feature vector into a 128-dimensional embedding, appends a learnable \texttt{[CLS]} token, and processes this sequence through a Transformer encoder (6 layers, 8 heads). The encoded \texttt{[CLS]} output is passed through a linear classifier to produce logits over the nine classes. To address class imbalance, we use a focal loss objective with class weights, defined as:
\[
\mathcal{L}= \frac{1}{B}\sum_{j=1}^{B}(1-p_j)^2 \cdot \text{CE}(\ell_j,y_j;\alpha),
\]
where $p_j=e^{-\text{CE}(\ell_j,y_j;\alpha)}$. The network is optimized with AdamW, and we apply early stopping based on the validation macro-F1 score.

\noindent\textbf{Integrated Gradients Attribution:} 
To identify which input features drive the Transformer’s decisions, we employ Integrated Gradients. Given an input $x$ and the zero-vector baseline $x'$, IG computes
\[
\mathrm{IG}_i(x) \;=\; (x_i - x'_i)\,\int_{0}^{1} \frac{\partial F\bigl(x' + \alpha(x - x')\bigr)}{\partial x_i}\,d\alpha,
\]
where $F$ denotes the trained model’s output score for the predicted class. We sample 100 test instances, compute each one's predicted class, and then calculate IG attributions with respect to that class. The resulting attribution vectors are aggregated by taking their mean absolute values to produce global importance rankings.

\noindent\textbf{Decision Tree Surrogate \& Rule Extraction:} 
We simplify the Transformer’s behavior into an interpretable decision tree. We train a CART decision tree (maximum depth=4, minimum leaf size=40) using the Transformer's predictions on the training data as pseudolabels. For each leaf node, we record the conjunction of feature-threshold conditions along the path from the root to form a logical clause:
\[
\texttt{class(}c\texttt{)} \;:-\; f_{i_1} \le \theta_1,\; f_{i_2} > \theta_2,\;\dots\;,
\]
where $c$ is the class predicted by that leaf. In a held-out test set, each instance belongs to exactly one leaf, we record the set of test indices covered by each leaf (the \emph{support set}). We then compute:
\begin{itemize}
    \item \emph{Coverage} = fraction of test instances assigned to any leaf in a given subset.
    \item \emph{Fidelity} = fraction of covered instances whose tree-predicted class matches the Transformer’s prediction.
    \item \emph{Redundancy} = mean pairwise Jaccard index among all leaf support sets.
\end{itemize}

\noindent\textbf{Baseline Comparisons:}
To contextualize the Transformer's performance, we compare it against a set of strong baseline models. This includes a Decision Tree, MLP, and powerful gradient boosting methods: \textbf{Random Forest, XGBoost, LightGBM, and CatBoost}. All baselines are trained on the same data and evaluated on the validation set. For LightGBM, we use \texttt{class\_weight='balanced'} to better handle class imbalance.

\subsection{LLM-Mediated Explanation and Validation:}
The framework culminates in generating and strictly validating natural language explanations. We employ a multi-LLM approach, using four different \textbf{generator models} (Qwen2.5:14b, llama3.1:8b, phi4:14b, gemma3:27b) to produce explanations. The prompt for the generator includes the predicted class, the triggered logical rule, and the top 5 most influential features from Integrated Gradients.

In our evaluation, we introduce a novel, three-part validation scheme to automatically assess the quality of each generated explanation:
\begin{enumerate}
    \item \textbf{Semantic Similarity:} Using a pre-trained \texttt{SentenceTransformer} model ('all-MiniLM-L6-v2'), we calculate the cosine similarity between the vector embeddings of the original logic rule and the LLM's text explanation. This measures how well the explanation preserves the core logical reasoning.
    \item \textbf{Attribution Faithfulness:} We programmatically check if the explanation's descriptive language (e.g., "high," "low," "increase") aligns with the mathematical sign (positive or negative) of the feature's attribution score. This ensures the explanation does not misrepresent the model's reasoning.
    \item \textbf{Actionability Score:} To assess the practical utility of an explanation for a security professional, we use a separate, powerful \textbf{evaluator LLM} (\texttt{llama3.3:70b}). The evaluator is prompted to score the explanation on a 1-5 scale of actionability, from "Not Actionable" to "Very Highly Actionable." This provides an expert-level judgment on the explanation's quality.
\end{enumerate}

\subsubsection{Prompt Design for Explanations and Evaluation}
To obtain consistent and high-quality outputs, we designed structured prompts for both the explanation generation and the actionability evaluation.

\paragraph{Generation Prompt}
For the generator LLMs, we use a constrained prompt template that encourages the model to align its wording with the feature attribution signs. For each instance, the prompt provides (i) the predicted class label, (ii) the logical rule (decision-tree clause) that fired, and (iii) the top-5 most influential features with their values and Integrated Gradients attribution scores. The instructions require the model to interpret positive attributions as "high," "large," or "key indicators" and negative attributions as "low," "small," or "not a concern." This reduces the likelihood of unsupported or contradictory statements and helps keep explanations grounded in the underlying attributions. The exact template used is shown in Figure~\ref{fig:gen_prompt}.

\begin{figure}[h!]
\centering
\footnotesize
\begin{tabular}{|p{0.95\columnwidth}|}
\hline
\ttfamily
\textbf{Generator Prompt Template}\\
\hline
You are a security analysis assistant. Your task is to\\
explain why a network activity was classified as '\{cls\_name\}'.\\
\\
Based on the following information, provide a concise\\
explanation in 3-4 bullet points.\\
\\
\textbf{Key Information:}\\
- Logical Rule Triggered: The activity matched the\\
\hspace*{1em}pattern: \{clause\}\\
- Top-5 Most Influential Features (with their actual\\
\hspace*{1em}values and attribution scores):\\
\{ig\_list\}\\
\\
\textbf{Instructions:}\\
- Return exactly 3-4 bullet points.\\
- Each bullet point must start with "- ".\\
- Each bullet point must reference one of the top-5\\
\hspace*{1em}features by its exact name (e.g., `\{sample\_feat\_name\}`).\\
- Each bullet point must discuss the feature's value\\
\hspace*{1em}and its influence.\\
- Crucially, use the sign of the 'Attribution' score to\\
\hspace*{1em}guide your language.\\
\hspace*{1em}- If a feature's Attribution is positive, describe\\
\hspace*{2em}its value as "high," "large," "elevated," or\\
\hspace*{2em}"a key indicator."\\
\hspace*{1em}- If a feature's Attribution is negative, describe\\
\hspace*{2em}its value as "low," "small," "absent," or\\
\hspace*{2em}"not a concern."\\
\\
\textbf{Example Response:}\\
- A high `ip.len` of 1400 bytes was a key indicator\\
\hspace*{1em}for this classification.\\
- The `tcp.time\_delta` was unusually low at 0.001s,\\
\hspace*{1em}strongly suggesting automated activity.\\
- The `tcp.flags` value of 18, while present, was\\
\hspace*{1em}noted as being less influential.\\
\hline
\end{tabular}
\caption{Structured prompt template used for generating explanations.}
\label{fig:gen_prompt}
\end{figure}

\paragraph{Evaluation Prompt}
To assess the practical utility of these explanations, we employ a separate, larger evaluator model (Llama3.3:70b) acting as an expert security analyst. The evaluator sees only the generated explanation text and is instructed to assign an actionability score on a 1--5 scale, where 1 denotes a non-actionable, generic description and 5 denotes a very highly actionable explanation that both justifies the alert and suggests a concrete next step or plausible attack type. The prompt emphasizes that high scores are reserved for explanations that provide security context rather than merely restating data values (see Figure~\ref{fig:eval_prompt}).

\begin{figure}[h!]
\centering
\footnotesize
\begin{tabular}{|p{0.95\columnwidth}|}
\hline
\ttfamily
\textbf{Evaluator Prompt Template}\\
\hline
You are an expert cybersecurity analyst. Your task is\\
to evaluate an explanation for a network security alert.\\
\\
\textbf{The Explanation to Evaluate:}\\
---\\
\{explanation\_text\}\\
---\\
\\
\textbf{Evaluation Criteria:}\\
On a scale of 1 to 5, how actionable is this\\
explanation for a security professional?\\
- 1 (Not Actionable): The explanation is generic,\\
\hspace*{1em}confusing, or simply repeats the data without\\
\hspace*{1em}context.\\
- 2 (Slightly Actionable): It mentions a feature\\
\hspace*{1em}value but provides no security context.\\
- 3 (Moderately Actionable): It correctly identifies\\
\hspace*{1em}a feature and links it to a general security\\
\hspace*{1em}concept (e.g., "high traffic").\\
- 4 (Highly Actionable): It explains \textit{why} a feature's\\
\hspace*{1em}value is suspicious, using relevant security\\
\hspace*{1em}terms (e.g., "This suggests a port scan").\\
- 5 (Very Highly Actionable): It not only explains\\
\hspace*{1em}the "why" but also suggests a potential attack\\
\hspace*{1em}type or a clear next step for investigation\\
\hspace*{1em}(e.g., "The pattern is consistent with a DDoS\\
\hspace*{1em}amplification attack").\\
\\
Based on these criteria, provide a score. Your\\
response must be only the score, in the format:\\
"Actionability Score: [score]"\\
\hline
\end{tabular}
\caption{Scoring rubric prompt used for the automated evaluator.}
\label{fig:eval_prompt}
\end{figure}

\section{Experimental Setup}
We evaluate the approach on a public 5G/IoT intrusion detection dataset collected on our testbed,  with separate training and testing splits. The dataset contains traffic data for nine classes (eight attacks and one benign). The framework is implemented in Python using PyTorch, scikit-learn, and the Hugging Face \texttt{transformers} library. All experiments were conducted on a single machine equipped with an NVIDIA RTX 4090 GPU, 128 GB of RAM, and an AMD Threadripper CPU. To ensure the robustness of the findings, all model training and evaluation experiments were repeated 5 times. A detailed summary of each train and test dataset is provided in Table~\ref{tab:dataset_summary}.

\begin{table}[h]
    \centering
    \scriptsize
    \caption{Summary of the Train and Test Datasets.}
    \label{tab:dataset_summary}
    \begin{tabular}{lrr}
        \toprule
        \textbf{Metric} & \textbf{Train Dataset} & \textbf{Test Dataset} \\
        \midrule
        Total Records & 1,753,454 & 194,829 \\
        Number of Features & 29 & 29 \\
        \addlinespace
        \multicolumn{3}{l}{\textbf{Class Distribution}} \\
        \midrule
        Benign & 1,322,254 & 148,070 \\
        Brute Force & 291 & 32 \\
        DDoS & 165,070 & 18,341 \\
        Device Spoofing & 70 & 8 \\
        DoS\_MQTT & 250,514 & 27,835 \\
        Eavesdropping & 3,525 & 392 \\
        MITM & 677 & 75 \\
        SQL Injection & 475 & 53 \\
        Unauthorized Data Access & 207 & 23 \\
        \bottomrule
    \end{tabular}
\end{table}

\subsection{Network Feature Set}
\label{subsec:feature_set}

We rely on 29 packet- and flow-level features spanning frame-, IP-, TCP-, UDP-, and HTTP-layer properties. 
Table~\ref{tab:network_features} summarizes all features and their descriptions.

\begin{table}[h!]
\centering
\caption{Network traffic features used in the model.}
\label{tab:network_features}
\footnotesize
\setlength{\tabcolsep}{30pt}
\begin{tabular}{ll}
\toprule
\textbf{Feature} & \textbf{Description} \\
\midrule
\multicolumn{2}{l}{\textit{HTTP Layer Features}} \\
\texttt{http.request.uri} & URI of HTTP request \\
\texttt{http.request} & Boolean flag for HTTP request \\
\midrule
\multicolumn{2}{l}{\textit{TCP Layer Features}} \\
\texttt{tcp.dstport} & Destination port number \\
\texttt{tcp.srcport} & Source port number \\
\texttt{tcp.port} & Source or destination port \\
\texttt{tcp.time\_delta} & Time since previous TCP segment \\
\texttt{tcp.time\_relative} & Time since first frame \\
\texttt{tcp.reassembled.length} & Total reassembled payload length \\
\texttt{tcp.segments} & Number of segments in PDU \\
\texttt{tcp.analysis.ack\_rtt} & Acknowledged round-trip time \\
\texttt{tcp.flags} & TCP flags bitmask \\
\texttt{tcp.urgent\_pointer} & TCP urgent pointer value \\
\texttt{tcp.stream} & Unique TCP stream identifier \\
\texttt{tcp.len} & TCP payload length (bytes) \\
\texttt{tcp.seq} & TCP sequence number \\
\texttt{tcp.ack} & TCP acknowledgment number \\
\texttt{tcp.ack\_raw} & Raw TCP acknowledgment \\
\texttt{tcp.window\_size.1} & TCP window size value \\
\midrule
\multicolumn{2}{l}{\textit{UDP Layer Features}} \\
\texttt{udp.port} & Source or destination port \\
\texttt{udp.length} & UDP datagram length (bytes) \\
\midrule
\multicolumn{2}{l}{\textit{IP Layer Features}} \\
\texttt{ip.proto} & Protocol number (e.g., 6=TCP) \\
\texttt{ip.ttl} & Time-to-live value \\
\texttt{ip.fragments} & Reassembled IP fragments \\
\texttt{ip.flags.mf} & More Fragments flag \\
\texttt{ip.flags.df} & Don't Fragment flag \\
\texttt{ip.len} & Total IP datagram length \\
\midrule
\multicolumn{2}{l}{\textit{Frame Layer Features}} \\
\texttt{frame.time\_delta} & Time since previous frame \\
\texttt{frame.time\_relative} & Time since first capture \\
\bottomrule
\end{tabular}
\end{table}

\subsection{Hyperparameter Tuning}
To select the optimal model configurations, we performed hyperparameter tuning based on performance on the validation set. For the main \textbf{Transformer IDS model}, we tuned the learning rate over $\{10^{-5}, 10^{-4}, 10^{-3}\}$ and the AdamW weight decay over $\{10^{-3}, 10^{-2}, 10^{-1}\}$. The final selected values were $10^{-4}$ and $10^{-2}$, respectively. For the \textbf{Decision Tree Surrogate}, we aimed to balance fidelity with interpretability. We performed a grid search for \texttt{max\_depth} over $\{3, 4, 5\}$ and \texttt{min\_samples\_leaf} over $\{20, 40, 60, 80\}$. The configuration of \texttt{max\_depth=4} and \texttt{min\_samples\_leaf=40} was chosen as it provided the best fidelity without creating an overly complex tree.

\subsection{LLM Configuration}
For all LLM-based explanation generation and evaluation tasks, we used a consistent set of parameters to ensure reproducibility and deterministic outputs. The API calls were configured with a \textbf{temperature of 0.1}. This low value was chosen to minimize randomness and ensure that the generated explanations are factual, consistent, and grounded in the provided prompt, which is essential for a scientific and analytical task. We also set a \textbf{max\_tokens limit of 250} to encourage concise, focused explanations suitable for a security analyst's review and to manage computational resources.

\subsubsection{Expert rating protocol}
In addition to the automatic, LLM-based evaluation of explanation quality, we also collected human judgements from two domain experts in network security and intrusion detection. Each expert interacted with a lightweight web-based survey interface that presented 20 explanation instances (five per generator LLM, covering both benign and attack traffic). For every instance, the experts rated the generated explanation along four 5-point Likert scales capturing structural validity, semantic consistency with the predicted class, perceived faithfulness to the described features, and practical actionability for a security analyst. These ratings are used only for evaluation (they are not fed back into the models) and provide an independent, human-centred view of explanation quality. A detailed analysis of the expert scores and their relationship to the LLM-as-judge metrics is reported in Section~\ref{subsec:expert-study}.

\section{Results and Analysis}
\begin{figure*}[h!]
\centering
\includegraphics[width=0.99\linewidth]{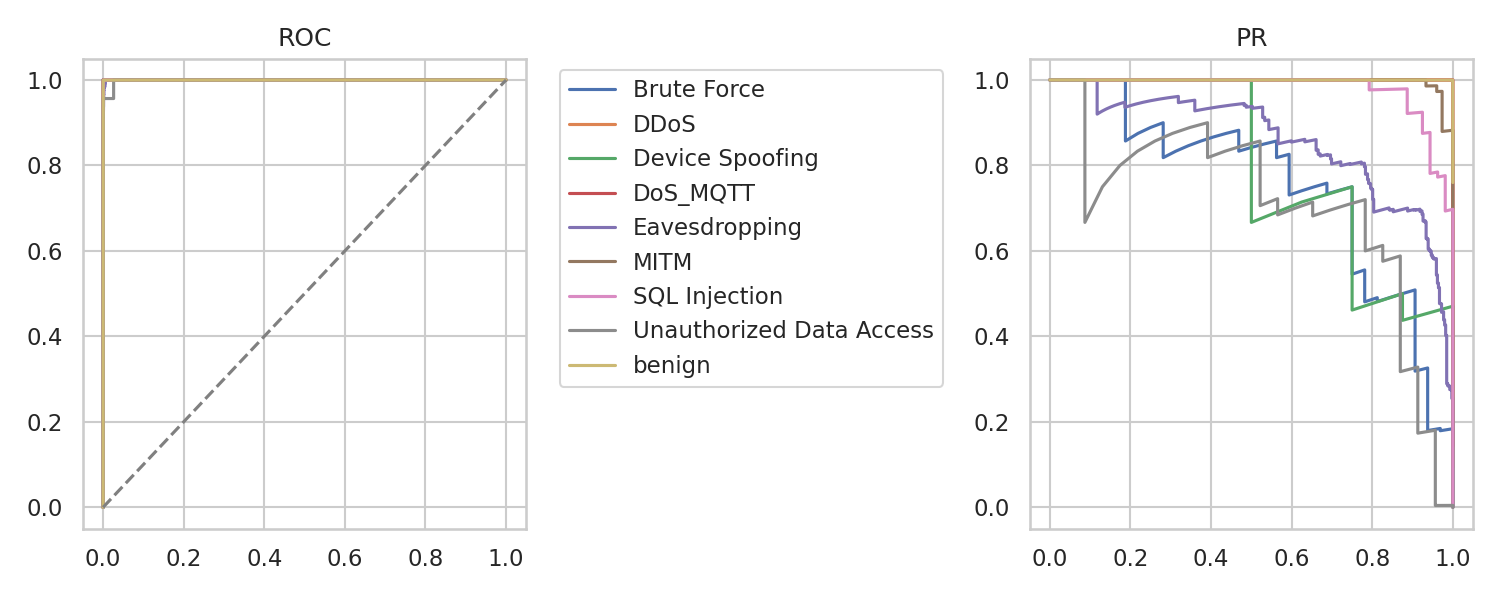}
\caption{ROC curves (left) and Precision-Recall curves (right) for all attack types.}
\label{fig:roc_pr_curves}
\end{figure*}

\subsection{Classification Performance}
The Transformer model achieves exceptional performance with an overall accuracy of \textbf{99.87\%} and a macro-averaged F1 score of \textbf{0.854}. Table~\ref{tab:classification_results} presents detailed per-class metrics that demonstrate robust performance across all attack categories despite severe class imbalance. Figure~\ref{fig:roc_pr_curves} shows the ROC and Precision-Recall curves for each attack type, illustrating the model's discriminative power. The curves demonstrate near-perfect performance for high-volume attack classes and strong performance for rare attack types.

\begin{table}[h]
\centering
\scriptsize
\caption{Per-class classification performance on test set.}
\label{tab:classification_results}
\begin{tabular}{lcccc}
\toprule
Attack Type & Precision & Recall & F1-Score & Support \\
\midrule
Brute Force & 0.619 & 0.812 & 0.703 & 32 \\
DDoS & 1.000 & 1.000 & 1.000 & 18341 \\
Device Spoofing & 0.714 & 0.625 & 0.667 & 8 \\
DoS\_MQTT & 1.000 & 1.000 & 1.000 & 27835 \\
Eavesdropping & 0.642 & 0.944 & 0.764 & 392 \\
MITM & 0.986 & 0.933 & 0.959 & 75 \\
SQL Injection & 0.978 & 0.830 & 0.898 & 53 \\
Unauthorized Data Access & 0.696 & 0.696 & 0.696 & 23 \\
Benign & 1.000 & 0.999 & 0.999 & 148070 \\
\midrule
\textbf{Macro Avg} & \textbf{0.848} & \textbf{0.871} & \textbf{0.854} & \textbf{194829} \\
\textbf{Weighted Avg} & \textbf{0.999} & \textbf{0.999} & \textbf{0.999} & \textbf{194829} \\
\bottomrule
\end{tabular}
\end{table}

\subsection{Baseline Comparison}
The performance of our explainable Transformer model relative to other baselines is shown in Table~\ref{tab:baseline_comparison}. While ensemble methods like RandomForest achieve a higher macro-F1 score, they operate as black boxes. The proposed framework provides an effective alternative, delivering high accuracy associated with good transparency. This result explicitly frames the critical trade-off for security domains: a marginal performance gain from an opaque model versus a highly performant and interpretable system.

\begin{table}[h]
\centering
\small
\caption{Baseline comparison on validation set (Macro-F1).}
\label{tab:baseline_comparison}
\begin{tabular}{lc}
\toprule
Model & Macro-F1 \\
\midrule
Decision Tree (depth 4) & 0.487 \\
MLP (256-128) & 0.887 \\
XGBoost & 0.966 \\
CatBoost & 0.970 \\
LightGBM & 0.980 \\
Random Forest & \textbf{0.989} \\
\midrule
\textbf{Transformer (Ours)} & 0.854 \\
\bottomrule
\end{tabular}
\end{table}

\subsection{Feature Attribution Analysis}
\begin{figure}[h]
\centering
\includegraphics[width=0.6\linewidth]{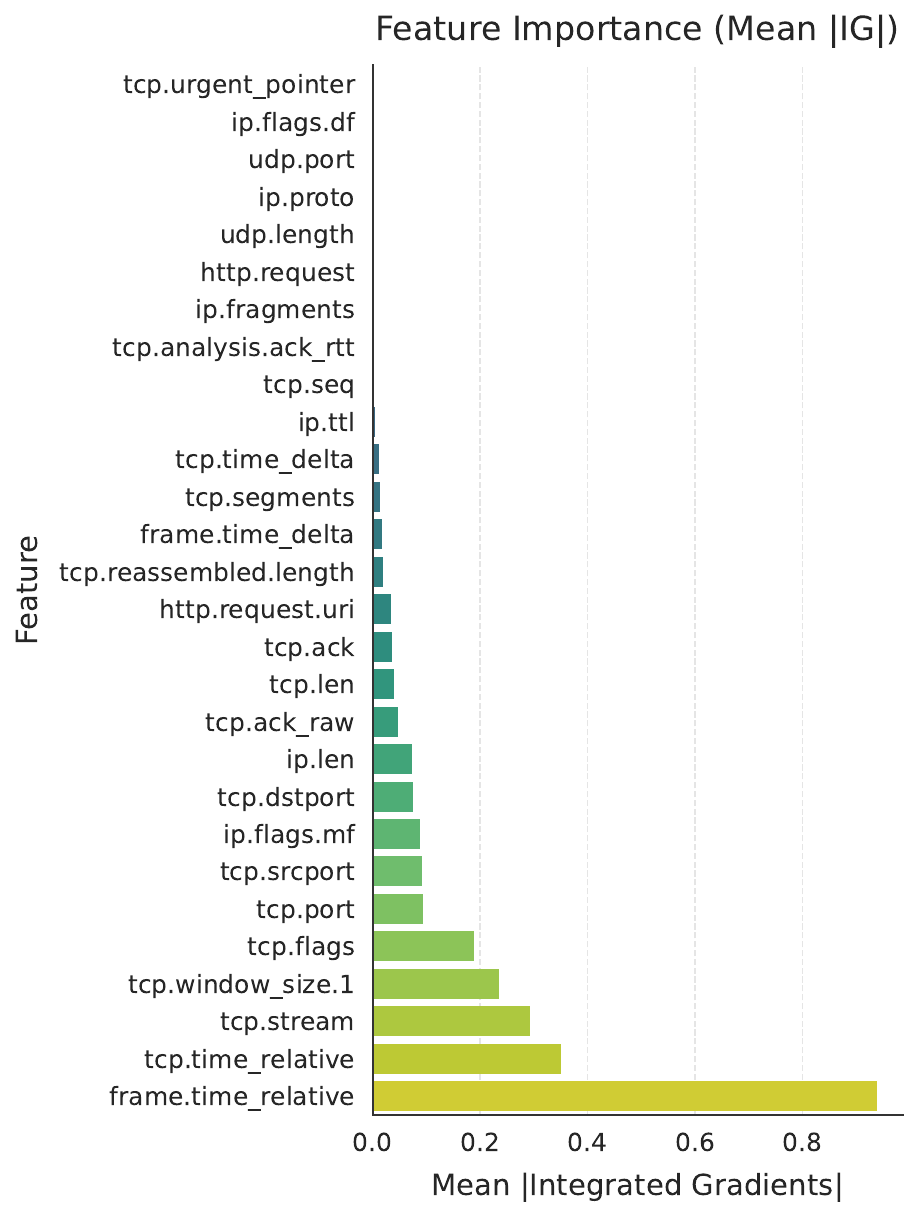}
\caption{Mean absolute Integrated Gradients attribution for top network features.}
\label{fig:feature_importance}
\end{figure}

Figure~\ref{fig:feature_importance} reveals that temporal and volumetric features dominate the model's decisions. \texttt{frame.time\_relative} has the highest mean attribution score, followed by \texttt{tcp.time\_relative} and \texttt{tcp.stream}. This aligns with domain knowledge that anomalies in packet timing and data flow size are key indicators of network intrusions.

\subsection{Rule Extraction and Interpretability}
\begin{figure}[h]
\centering
\includegraphics[width=\linewidth]{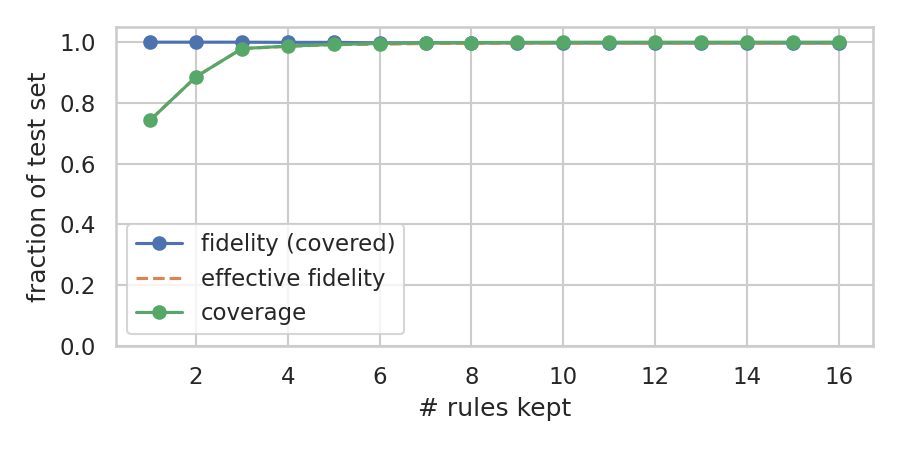}
\caption{Coverage and fidelity trade-off as rules are added in descending support order.}
\label{fig:coverage_fidelity}
\end{figure}

The surrogate decision tree generates \textbf{16} interpretable rules with \textbf{99.72\% ± 0.01\%} fidelity and \textbf{100\%} coverage across all runs. After pruning the single least-supported leaf, the remaining 15 rules maintain their high fidelity while still covering 99.998\% of test cases. Figure~\ref{fig:coverage_fidelity} demonstrates that the eight highest-support rules are sufficient to explain over 99\% of instances, enabling efficient rule-based deployment and analysis.

\subsection{Multi-LLM Explanation Quality and Validation}
We evaluated explanations from four LLMs across a more robust set of \textbf{20 randomly sampled test instances}. The aggregated results, shown in Table~\ref{tab:llm_evaluation}, reveal a significant difference in quality among models. While all models produced structurally valid outputs, their ability to generate faithful and actionable content varied.

\textbf{Qwen2.5:14b} and \textbf{phi4:14b} emerge as top performers, achieving near-perfect attribution faithfulness scores of \textbf{0.99} and \textbf{0.98} respectively. This indicates they correctly interpret and represent the direction of influence (positive or negative) for each feature. Furthermore, both models achieved excellent average actionability scores of \textbf{4.6} and \textbf{4.5}, demonstrating their ability to produce explanations that are highly useful to a security professional. In contrast, gemma3:27b had a lower faithfulness score (0.91), suggesting it sometimes struggles to correctly map attribution signs to descriptive language. This more extensive evaluation confirms that with the right model and prompting, it is possible to generate AI explanations that are consistently trustworthy and useful.

\begin{table}[ht!]
\centering
\small
\setlength{\tabcolsep}{3pt}        
\renewcommand{\arraystretch}{1.15} 
\caption{LLM Explanation Evaluation Results (Averaged over 20 test instances). Scores for Faithfulness range from 0 to 1. Actionability is scored on a scale of 1--5.}
\label{tab:llm_evaluation}

\begin{tabularx}{\columnwidth}{lCCCC}
\toprule
\textbf{Generator} &
\makecell{\textbf{Struct.}\\\textbf{valid (\%)}} &
\makecell{\textbf{Semantic}\\\textbf{similarity}} &
\makecell{\textbf{Attribution}\\\textbf{faithfulness}} &
\makecell{\textbf{Actionability}\\\textbf{(1--5)}} \\
\midrule
gemma3\_27b    & 100\% & 0.670 & 0.91         & 3.9 \\
llama3.1\_8b   & 100\% & 0.668 & 0.94         & 4.1 \\
phi4\_14b      & 100\% & 0.678 & \textbf{0.98} & 4.5 \\
Qwen2.5\_14b   & 100\% & 0.678 & \textbf{0.99} & \textbf{4.6} \\
\bottomrule
\end{tabularx}
\end{table}

An example of a high-quality explanation for a \texttt{DoS\_MQTT} attack (ID: 60492) from \textbf{phi4:14b} highlights these strengths:
\begin{itemize}
    \small
    \item - A \textbf{high} `frame.time\_relative` of 812.4183 is a key indicator, contributing significantly to the classification of DoS\_MQTT due to its high attribution score.
    \item - The `tcp.time\_relative` value of 0.0000 is "small", aligning with expected values for this pattern and strongly supporting the DoS\_MQTT classification due to its negative attribution.
    \item - A "large" `tcp.stream` count of 598269.0000 serves as a crucial factor, positively influencing the identification of DoS\_MQTT activity based on its notable attribution score.
\end{itemize}
This explanation is effective because it correctly uses directional language ("high," "small," "large"), aligns perfectly with the feature attribution signs, and provides valuable security context (e.g., "high volume of simultaneous connections," "sudden and rapid packet transmissions").

\subsubsection{Qualitative Example Explanations}
To complement the quantitative metrics, Table~\ref{tab:sample_explanations} presents the actual generated text from all four LLMs for two distinct scenarios: a DoS\_MQTT attack and a Benign traffic instance. The table highlights the difference in semantic quality; while all models generate coherent text, the top performers (Phi4 and Qwen) use more precise directional language.

An example of a high-quality explanation for a \texttt{DoS\_MQTT} attack (ID: 60492) from \textbf{phi4:14b} highlights these strengths:
\begin{itemize}
    \small
    \item - A \textbf{high} `frame.time\_relative` of 812.4183 is a key indicator, contributing significantly to the classification of DoS\_MQTT due to its high attribution score.
    \item - The `tcp.time\_relative` value of 0.0000 is "small", aligning with expected values for this pattern and strongly supporting the DoS\_MQTT classification due to its negative attribution.
    \item - A "large" `tcp.stream` count of 598269.0000 serves as a crucial factor, positively influencing the identification of DoS\_MQTT activity based on its notable attribution score.
\end{itemize}
This explanation is effective because it correctly uses directional language ("high," "small," "large"), aligns perfectly with the feature attribution signs, and provides valuable security context (e.g., "high volume of simultaneous connections," "sudden and rapid packet transmissions").

\begin{table}[H]
\centering
\scriptsize
\caption{Representative LLM-generated explanations for DoS\_MQTT and benign traffic.}
\label{tab:sample_explanations}
\setlength{\tabcolsep}{4pt}
\renewcommand{\arraystretch}{1.2}
\begin{tabularx}{\textwidth}{@{}l>{\raggedright\arraybackslash}X@{}}
\toprule
\textbf{Model} & \textbf{Generated Explanation} \\
\midrule
\multicolumn{2}{c}{\textit{\textbf{Record 60492: DoS\_MQTT Attack}}} \\
\midrule
\textbf{Gemma3:27b} & 
\vspace{-0.3em}
\begin{itemize}[leftmargin=*, topsep=0pt, partopsep=0pt, parsep=0pt, itemsep=1pt]
\item A \textbf{high} \texttt{frame.time\_relative} of 812.4183 was a key indicator, contributing significantly to the DoS classification.
\item The \texttt{tcp.time\_relative} value of 0.0000 was \textbf{low} and therefore not a primary driver of this classification.
\item A \textbf{large} \texttt{tcp.stream} value of 598269.0000 suggests a high volume of connections, supporting the DoS categorization.
\item The \texttt{tcp.window\_size.1} of 64.0000 was \textbf{elevated}, contributing to the overall assessment of potentially malicious activity.
\end{itemize}
\vspace{-0.3em} \\
\textbf{Llama3.1:8b} & 
\vspace{-0.3em}
\begin{itemize}[leftmargin=*, topsep=0pt, partopsep=0pt, parsep=0pt, itemsep=1pt]
\item A \textbf{large} \texttt{frame.time\_relative} of 812.4183 seconds was a key indicator for this classification, indicating that the network activity occurred far into the past.
\item The \texttt{tcp.stream} value of 598269.0000 was \textbf{high}, suggesting an elevated stream ID, which contributed to the DoS\_MQTT classification.
\item A \textbf{low} \texttt{tcp.time\_relative} of 0.0000 seconds was not a concern in this context.
\end{itemize}
\vspace{-0.3em} \\
\textbf{Phi4:14b} & 
\vspace{-0.3em}
\begin{itemize}[leftmargin=*, topsep=0pt, partopsep=0pt, parsep=0pt, itemsep=1pt]
\item A \textbf{high} \texttt{frame.time\_relative} of 812.4183 is a key indicator, contributing significantly to the classification of DoS\_MQTT due to its high attribution score.
\item The \texttt{tcp.time\_relative} value of 0.0000 is \textbf{small}, aligning with expected values for this pattern and strongly supporting the DoS\_MQTT classification due to its negative attribution.
\item A \textbf{large} \texttt{tcp.stream} count of 598269.0000 serves as a crucial factor, positively influencing the identification of DoS\_MQTT activity.
\end{itemize}
\vspace{-0.3em} \\
\textbf{Qwen2.5:14b} & 
\vspace{-0.3em}
\begin{itemize}[leftmargin=*, topsep=0pt, partopsep=0pt, parsep=0pt, itemsep=1pt]
\item A \textbf{high} \texttt{frame.time\_relative} value of 812.4183 seconds was a key indicator for this classification.
\item The \texttt{tcp.time\_relative} value of 0.0000 seconds was \textbf{low}, not a concern in terms of influencing the decision.
\item The \texttt{tcp.window\_size.1} value of 64 bytes was large, elevating its influence on the classification.
\end{itemize}
\vspace{-0.3em} \\
\midrule
\multicolumn{2}{c}{\textit{\textbf{Record 139096: Benign Traffic}}} \\
\midrule
\textbf{Gemma3:27b} & 
\vspace{-0.3em}
\begin{itemize}[leftmargin=*, topsep=0pt, partopsep=0pt, parsep=0pt, itemsep=1pt]
\item A \textbf{low} \texttt{tcp.time\_relative} of 519.3665 contributed to the benign classification, as indicated by its negative attribution score.
\item The \texttt{tcp.port} value of 44350.0000 was \textbf{a key indicator} of benign activity, supported by its positive attribution score.
\item The \texttt{tcp.flags} value of 24.0000 was \textbf{not a concern} for malicious activity, given its slightly negative attribution score.
\item An \textbf{elevated} \texttt{tcp.stream} value of 94.0000 played a role in classifying this activity as benign.
\end{itemize}
\vspace{-0.3em} \\
\textbf{Llama3.1:8b} & 
\vspace{-0.3em}
\begin{itemize}[leftmargin=*, topsep=0pt, partopsep=0pt, parsep=0pt, itemsep=1pt]
\item A \textbf{high} \texttt{tcp.time\_relative} value of 519.3665 indicates that the network activity is likely to be benign due to its prolonged duration.
\item The \texttt{tcp.port} of 44350 was a \textbf{large} contributor to this classification, indicating a potential legitimate service or application.
\item The \textbf{absent} influence of \texttt{tcp.flags}, with an attribution score of -0.093, suggests that the network activity does not exhibit suspicious flag settings.
\end{itemize}
\vspace{-0.3em} \\
\textbf{Phi4:14b} & 
\vspace{-0.3em}
\begin{itemize}[leftmargin=*, topsep=0pt, partopsep=0pt, parsep=0pt, itemsep=1pt]
\item The \texttt{tcp.time\_relative} value of 519.3665 is considered \textbf{low}, indicating that the timing of this packet relative to others in the session does not raise concern and aligns with benign behavior.
\item A \textbf{high} \texttt{tcp.port} of 44350.0000 serves as a key indicator for classification, suggesting typical usage patterns such as secure web traffic.
\item The \texttt{tcp.flags} value of 24.0000 was noted to be \textbf{less influential}, reinforcing the activity's non-threatening nature.
\end{itemize}
\vspace{-0.3em} \\
\textbf{Qwen2.5:14b} & 
\vspace{-0.3em}
\begin{itemize}[leftmargin=*, topsep=0pt, partopsep=0pt, parsep=0pt, itemsep=1pt]
\item The \texttt{tcp.time\_relative} value of 519.3665 seconds was considered \textbf{low}, indicating minimal concern for this classification.
\item A \texttt{tcp.port} value of 44350 was considered a standard port assignment, contributing positively to the benign assessment.
\item The \texttt{tcp.flags} value of 24 had a minor influence as it was \textbf{less critical} in determining the activity's nature.
\end{itemize}
\vspace{-0.3em} \\
\bottomrule
\end{tabularx}
\end{table}

\subsection{Expert Study of Explanation Quality}
\label{subsec:expert-study}

To complement the automatic evaluation of explanation quality in Section~5.5, we conducted a small expert study to obtain a human-centered assessment of the generated explanations. The goal was to understand how security professionals perceive different generator LLMs along dimensions closely aligned with our automatic metrics (semantic similarity, attribution faithfulness, and actionability), and to test whether the LLM-as-judge scores approximate expert judgements.

\paragraph{Participants and task}
We collected responses from two domain experts in network security and intrusion detection. The first expert has over ten years of professional experience, and the second expert has between five and ten years of experience. Both experts interacted with the same web-based interface, which, for each survey item, presented the record identifier, the predicted class (\emph{benign}, \emph{DDoS}, or \emph{DoS\_MQTT}), the generator LLM, and the corresponding natural language explanation.

For each explanation, the experts provided ratings on four 5-point Likert scales (1~{=}\ very poor, 5~{=}\ very high):

\begin{itemize}
    \item \textbf{Structural validity} (m1): how well-formed, coherent, and readable the explanation is.
    \item \textbf{Semantic consistency} (m2): how well the explanation matches the stated class and the described traffic pattern.
    \item \textbf{Faithfulness} (m3): how well the explanation appears to focus on the salient features it mentions, without introducing irrelevant or contradictory factors.
    \item \textbf{Actionability} (m4): how useful the explanation would be for a security analyst deciding whether and how to respond.
\end{itemize}

These four dimensions mirror the goals of the automatic evaluation: semantic similarity (\emph{semantic consistency}), attribution faithfulness (\emph{faithfulness}), and evaluator-LLM actionability (\emph{actionability}), while also capturing general linguistic quality (\emph{structural validity}).

\paragraph{Materials}
The study utilized a total of 20 explanation instances. For each of the four generator models considered in Section~5.5 (Qwen2.5:14b, Gemma3:27b, Llama3.1:8b, Phi4:14b), we sampled five explanations covering both benign and attack traffic (\emph{DDoS} and \emph{DoS\_MQTT}). Each item, therefore, corresponds to a unique combination of record, predicted class, and generator model. Both experts rated all 20 items, resulting in 40 explanation--annotator pairs and 160 scalar ratings in total (20 items $\times$ 2 experts $\times$ 4 metrics).

\paragraph{Overall results}
Averaged over all explanations and both experts, the mean scores on the 1--5 scale are:
\begin{itemize}
    \item Structural validity (m1): \textbf{4.0}
    \item Semantic consistency (m2): \textbf{3.6}
    \item Faithfulness (m3): \textbf{3.7}
    \item Actionability (m4): \textbf{3.6}
\end{itemize}
These moderate-to-high ratings indicate that, from a two-expert perspective, the LLM-generated explanations are generally perceived as understandable and reasonably useful, but still leave room for improvement, particularly in faithfully reflecting the underlying model behaviour and in providing highly actionable guidance.

\paragraph{Per-model analysis and comparison with LLM-based metrics}
Table~\ref{tab:expert-vs-llm} compares the experts' average ratings per generator model to the automatic metrics reported in Section~5.5 for the same set of models. For each generator LLM, we report the mean structural validity, semantic consistency, faithfulness, and actionability across both experts (all 1-5), alongside the semantic similarity, attribution faithfulness (0-1), and evaluator-LLM actionability score (1-5) used in the LLM-as-judge evaluation.

\begin{table}[!t]
\centering
\scriptsize
\setlength{\tabcolsep}{3pt}
\renewcommand{\arraystretch}{1.15}

\caption{Comparison of two-expert ratings (1-5) with automatic LLM-based metrics from the multi-LLM evaluation. Human ratings are averaged over five explanations per generator model, rated independently by two experts (10 ratings per model and metric). Semantic similarity and attribution faithfulness range from 0 to 1, while actionability is on a 1-5 scale.}
\label{tab:expert-vs-llm}

\begin{tabularx}{\columnwidth}{@{}lCCCCCCC@{}}
\toprule
& \multicolumn{4}{c}{Human experts (1--5)} & \multicolumn{3}{c}{Automatic LLM-based metrics} \\
\cmidrule(lr){2-5}\cmidrule(lr){6-8}
\makecell{\textbf{Generator}\\\textbf{LLM}} &
\textbf{Struct.} & \textbf{Sem.} & \textbf{Faith.} & \textbf{Act.} &
\makecell{\textbf{Sem.}\\\textbf{sim.}} &
\makecell{\textbf{Attr.}\\\textbf{faith.}} &
\makecell{\textbf{Act.}\\\textbf{(1--5)}} \\
\midrule
Gemma3:27b  & 4.4 & 3.4 & 3.3 & 3.4 & 0.670 & 0.91 & 3.9 \\
Llama3.1:8b & 3.4 & 3.5 & 3.3 & 3.3 & 0.668 & 0.94 & 4.1 \\
Phi4:14b    & 4.4 & 3.7 & 4.1 & 4.1 & 0.678 & 0.98 & 4.5 \\
Qwen2.5:14b & 3.8 & 3.7 & 3.9 & 3.5 & 0.678 & 0.99 & 4.6 \\
\bottomrule
\end{tabularx}
\end{table}

The experts consistently rate all models as structurally sound (scores between 3.4 and 4.4), which is in line with the automatic structural validity metric reporting 100\% valid outputs for all four LLMs. Phi4:14b and Gemma3:27b obtain the highest structural validity scores (4.4), while Qwen2.5:14b and Llama3.1:8b are slightly lower but still comfortably above the midpoint.

For semantic consistency and faithfulness, the human ratings are more differentiated than the automatic scores. Phi4:14b achieves the highest faithfulness (4.1) and also strong semantic consistency (3.7), followed closely by Qwen2.5:14b (3.9 faithfulness, 3.7 semantic consistency). Gemma3:27b and Llama3.1:8b obtain more moderate scores around 3.3--3.5. Interestingly, despite all four models achieving high automatic attribution faithfulness (0.91-0.99), the experts perceive more variation, suggesting that faithfulness, as inferred from the attribution signal, and faithfulness, as perceived from the explanation text, are related but not identical.

Regarding actionability, the evaluator LLM produces relatively high scores for all models (3.9-4.6), whereas the human experts are more conservative, with ratings ranging from 3.3 to 4.1. Phi4:14b again stands out with the highest human actionability score (4.1), broadly consistent with its strong automatic actionability score. The other models cluster closely behind, with Qwen2.5:14b at 3.5 and Gemma3:27b and Llama3.1:8b at 3.4 and 3.3, respectively.

Although this study still involves a small sample of two experts and 20 explanation instances, it provides a useful sanity check and qualitative complement to the LLM-as-judge metrics. First, all four LLMs produce explanations that are judged as at least moderately coherent, semantically aligned, and actionable by practitioners. Second, the comparison in Table~\ref{tab:expert-vs-llm} highlights that automatic scores tend to be slightly more optimistic and that model rankings can differ between human and LLM evaluators, especially for faithfulness and semantic alignment. This highlights the importance of retaining a human-in-the-loop perspective when deploying explanation systems in critical 5G security settings. Future work will extend this study to more analysts and a larger set of explanations to quantify inter-rater agreement and to further probe the alignment between automated and human-centred evaluation of explanation quality.

\subsection{Computational Efficiency}
The system achieves excellent efficiency with a median inference latency of \textbf{2.48 ms} per flow on a CPU. This high throughput enables real-time deployment in demanding 5G network environments without requiring specialized hardware.

\begin{figure}[ht!]
\centering
\includegraphics[width=0.5\linewidth]{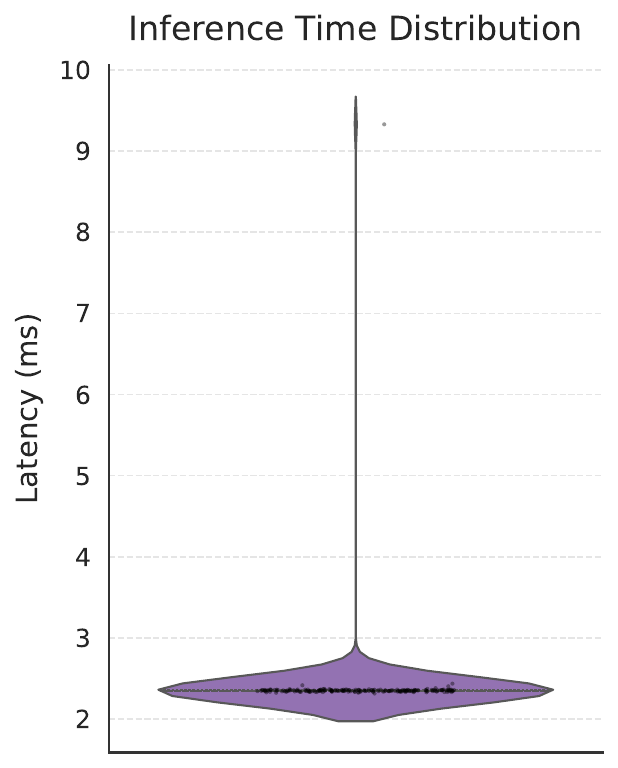}
\caption{Inference latency distribution for the Transformer model (200 test samples).}
\label{fig:latency_dist}
\end{figure}

\section{Discussion}
The results prompt a critical discussion on AI performance in cybersecurity. The baseline comparison clearly shows that unclear ensemble models, such as RandomForest, can achieve a higher macro-F1 score than the Transformer-based model. A purely academic, metric-driven perspective might interpret this as a weakness. However, from an operational standpoint, this result is the central strength and argument of the work.

In a security operations center, an analyst faced with an alert from a black-box model confronts a dilemma: trust the alert without understanding its basis or ignore it, risking a breach. This trust deficiency can make even the highest-performing models operationally useless. The ExAI5G framework confronts this problem directly. We demonstrate that it is possible to build a system that is both highly accurate (99.9\%) and completely transparent. The slight difference in F1-score is a necessary trade-off for achieving full interpretability, which is arguably essential for systems deployed in critical 5G infrastructure.

The real success of the proposed framework is not simply about surpassing the baseline F1-score; it's about offering a reliable and trustworthy alternative. Every decision made by this system can be deconstructed into a simple logical rule, grounded in feature attributions, and expressed in natural language that we have quantitatively proven to be faithful and actionable. This shifts the discussion from narrowly optimizing metrics to a more advanced conversation about creating AI systems that are robust, reliable, and worthy of human trust.

\section{Limitations and Future Work}
While this framework successfully integrates performance and interpretability, we recognize several limitations that provide opportunities for future research.

\noindent\textbf{Limitations:}
\begin{itemize}
    \item \textbf{Dataset Scope:} The evaluation is conducted on a single, collected comprehensive, 5G/IoT dataset. The performance and the extracted rules may not generalize perfectly to different network environments.
    \item \textbf{Surrogate Model Fidelity:} While the surrogate decision tree achieves high fidelity (99.72\%), it is not a perfect one-to-one mapping of the Transformer model. A small fraction of the deep model's decisions are not captured by the logic rules, indicating a remaining area of opacity.
    \item \textbf{Adversarial Robustness:} The very transparency of the system could be a vector for attack. An adversary with knowledge of the extracted rules could potentially craft malicious traffic specifically designed to evade detection. The adversarial robustness of this explanation-driven framework has not yet been evaluated.
\end{itemize}

\noindent\textbf{Future Work:}
Building on these limitations, we propose several directions for future work:
\begin{itemize}
    \item \textbf{Cross-Dataset Validation:} To ensure generalizability, the ExAI5G framework should be tested on a wider variety of datasets from different 5G core network vendors and IoT deployment scenarios.
    \item \textbf{Enhancing Explanation Faithfulness:} Future research could explore more complex surrogate models, such as rule ensembles or programmatic rule induction techniques, to close the fidelity gap while maintaining interpretability.
    \item \textbf{Adversarial Defense for XAI:} A critical next step is to investigate the system's vulnerability to explanation-aware adversarial attacks and to develop corresponding defense mechanisms, potentially by introducing a degree of randomness or ensemble-based explanations.
    \item \textbf{Automated Rule-to-Response Pipelines:} The high-fidelity logical rules generated by the system are well-suited for automated security orchestration. Future work could focus on building pipelines that automatically translate these rules into active defense policies, such as generating firewall rules or SIEM correlation searches.
\end{itemize}


\section{Conclusion}
This work introduced ExAI5G, an explainable AI framework designed to bring transparency to intrusion detection in 5G networks. We have demonstrated that it is not necessary to accept the opacity of high-performing models as an unavoidable trade-off. The proposed Transformer-based system achieves 99.9\% accuracy and extracts a set of 16 logical rules that represent its decision-making process with 99.7\% fidelity.

The central contribution of this paper is not the creation of another high-scoring IDS, but the presentation of a holistic, trustworthy system. The use of prioritizing explainability in our method directly addresses the critical gap between AI models and the human operators who must rely on them. The novel, quantitative validation of LLM-generated explanations further ensures that this transparency is meaningful and actionable. Finally, ExAI5G proves that the most effective IDS for a critical domain is not necessarily the one with the highest F1-score, but the one whose performance is matched by its ability to be understood, verified, and trusted.

\section*{Acknowledgements}

This work was supported by the Research Council of Finland through the 6G Flagship program (grant 318927), the Strategic Research Council affiliated with Academy of Finland through the CO2CREATION project (grant 372355), by Business Finland through the Neural pub/sub research project (diary number 8754/31/2022), and by the ERDF (project numbers A81568, A91867).

\section*{CRediT authorship contribution statement}

Saeid Sheikhi: Conceptualization, Methodology, Software, Data curation, Formal analysis, Investigation, Visualization, Writing, original draft.

Panos Kostakos: Supervision, Project administration, Validation, Writing, review \& editing.

Lauri Lovén:  Supervision, Validation, Writing, review \& editing, Funding acquisition.

\section*{Ethics statement}

The network traffic dataset used in this study was collected in a fully controlled 5G testbed without real user traffic or personally identifiable information. All attack scenarios were synthetically generated for research purposes.

The expert evaluation in Section~5.6 involved two adult security professionals who participated voluntarily and provided informed consent. No personally identifiable or sensitive information about the experts was collected, and the ratings are reported only in aggregate form.

\section*{Declaration of competing interest}
The authors declare that they have no conflicts of interest about this study.

\bibliographystyle{elsarticle-num} 
\bibliography{bib}
\end{document}